\documentclass[english,reprint]{revtex4-1}
\usepackage[T1]{fontenc}
\usepackage[latin9]{inputenc}
\usepackage{geometry}
\usepackage{babel}
\begin{document}
\title{On contextuality and logic in scientific discourse: Reply to Aliakbarzadeh,
Kitto, and Bruza}
\author{Ehtibar N.\ Dzhafarov}
\email{E-mail: ehtibar@purdue.edu}

\affiliation{Purdue University, USA}
\begin{abstract}
Aliakbarzadeh, Kitto, and Bruza (arXiv:1909.13048v1) criticize the
Contextuality-by-Default (CbD) theory in three ways: (1) they claim
an internal contradiction within the theory, that consists in both
denying and accepting the equality of stochastically unrelated random
variables; (2) they find CbD deficient because when one confines one's
attention to consistently connected systems (those with no disturbance/signaling)
one can dispense with the double-indexation of random variables; and
(3) they see no relationship between double-indexation and ontic states.
This commentary shows that (re$\:$1) the contradiction is inserted
in the presentation of CbD by the authors of the paper; (re$\:$2)
in the case of consistently connected systems CbD properly specializes
to traditional understanding of contextuality; and (re$\:$3) if ontic
states are understood as quantum states, they are simply not within
the language of CbD, and if they are understood as domain probability
spaces for context-sharing random variables, they can be trivially
reconstructed.
\end{abstract}
\maketitle
Aliakbarzadeh, Kitto, and Bruza, in Ref. \citep{AKB}, henceforth
referred to as {[}AKB{]}, criticize the Contextuality-by-Default (CbD)
theory \citep{DzhCerKuj2017,DzhKujFoundations2017,KujDzhMeasures}
using three methods of argumentation, summarized in the three subsequent
section titles.

\subsection*{Method 1: Insert a contradiction, then critisize it}

The main criticism of CbD in {[}AKB{]} is based on the analysis of
Ernst Specker's parable of the seer with three magic boxes \citep{LiangSpekkensWiseman2011}.
Mathematically, it is a special case of a cyclic system of rank 3
\citep{KujDzhLar2015,DzhKujLar2015}: in CbD notation,

\begin{equation}
\begin{array}{|c|c|c||c|}
\hline a_{1}^{1} & a_{2}^{1} &  & c^{1}\\
\hline  & a_{2}^{2} & a_{3}^{2} & c^{2}\\
\hline a_{1}^{3} &  & a_{3}^{3} & c^{3}\\
\hline\hline q_{1} & q_{2} & q_{3} & \mathcal{C}_{3}
\\\hline \end{array},\label{eq:C3}
\end{equation}
where $a_{i}^{j}$ is a dichotomous (say, $\pm1$) random variable
measuring property $q_{i}$ in context $c_{j}$. Specker's special
case of this system is a PR box with uniform marginals: $\Pr\left[a_{i}^{j}=1\right]=0.5$,
for all $i,j$, and 
\begin{equation}
\Pr\left[a_{1}^{1}=a_{2}^{1}\right]=\Pr\left[a_{2}^{2}=a_{3}^{2}\right]=\Pr\left[a_{1}^{3}=-a_{3}^{3}\right]=1.
\end{equation}
{[}AKB{]} correctly points out that in CbD any two $a_{i}^{j}$ and
$a_{i}^{j'}$ are \emph{stochastically unrelated} (do not have a joint
distribution) if $j\not=j'$. It follows that no probability like
$\Pr\left[a_{i}^{j}=a_{i}^{j'}\right]$ is defined for such random
variables. This expression is meaningless.

However, when {[}AKB{]} provides a definition of \emph{consistent
connectedness}, central for the paper, the authors get more creative.
The first part of Definition 2 (p. 3) states that a system is consistently
connected if $a_{i}^{j}\sim a_{i}^{j'}$ (i.e., the variables are
identically distributed) for any $i,j,j'$ for which the random variables
exist. This is a correct presentation of the CbD definition: e.g.,
in (\ref{eq:C3}), $a_{2}^{1}\sim a_{2}^{2}$, $a_{3}^{2}\sim a_{3}^{3}$,
and $a_{1}^{3}\sim a_{1}^{1}$ (because they are all uniformly distributed).
The second part of Definition 2, however, is where the creative insertion
occurs:
\begin{quote}
\begin{small}Alternatively, this relation is denoted $\Pr\left[a_{i}^{j}=a_{i}^{j'}\right]=1$
(p. 3).\end{small}
\end{quote}
No, the authors of {[}AKB{]} should be told, the relation $a_{i}^{j}\sim a_{i}^{j'}$
is not ``denoted'' so, alternatively or otherwise: $\Pr\left[a_{i}^{j}=a_{i}^{j'}\right]$
is undefined because $a_{i}^{j}$ and $a_{i}^{j'}$ are not jointly
distributed. This logical error is inserted immediately after citing
a paper by de Barros, Kujala, and Oas \citep{deBarrosetal.2016},
so that a reader not familiar with the issues should think it was
taken from there.

The correct way system $\mathcal{C}_{3}$ in (\ref{eq:C3}) in analyzed
in CbD is as follows (see, e.g., \citep{DzhKujLar2015}). We consider
all possible \emph{couplings} $S$ of $\mathcal{C}_{3}$, i.e., all
sextuples of jointly distributed random variables 
\begin{equation}
\begin{array}{|c|c|c||c|}
\hline S_{1}^{1} & S_{2}^{1} &  & c^{1}\\
\hline  & S_{2}^{2} & S_{3}^{2} & c^{2}\\
\hline S_{1}^{3} &  & S_{3}^{3} & c^{3}\\
\hline\hline q_{1} & q_{2} & q_{3} & S
\\\hline \end{array},\label{eq:coupling}
\end{equation}
that have the same row-wise distributions as $\mathcal{C}_{3}$, and
we ask whether this class of couplings contains one which is \emph{maximally
connected}. The latter means that the probability of $S_{i}^{j}=S_{i}^{j'}$
is maximal possible for all $i,j,j'$ for which the random variables
exist. A simple theorem tells us that since $S_{i}^{j}\sim S_{i}^{j'}$,
the maximal possible $\Pr\left[S_{i}^{j}=S_{i}^{j'}\right]$ is 1.
For system $\mathcal{C}_{3}$ such a coupling does not exist, and
this means that $\mathcal{C}_{3}$ is a \emph{contextual system}.
The non-existence of such a coupling can be established in various
ways, e.g., by invoking the general theory of cyclic systems \citep{KujDzhProof2016,Araujoetal2013}
and showing that $\mathcal{C}_{3}$ violates the corresponding Bell-type
inequality, presented below in (\ref{eq:SZLG}). In this especially
simple case, however, it can also be shown by \emph{reductio}: if
such a coupling exists, then with probability 1 we should have
\begin{equation}
S_{1}^{1}=S_{2}^{1}=S_{2}^{2}=S_{3}^{2}=S_{3}^{3}=-S_{1}^{3}=-S_{1}^{1},
\end{equation}
which is impossible since $S_{1}^{1}$ is not zero.

The authors of {[}AKB{]} begin their analysis of CbD by providing
a version of this \emph{reductio} reasoning. However, they apply it
to system $\mathcal{C}_{3}$ itself, using $a_{i}^{j}$'s in place
of the jointly distributed $S_{i}^{j}$'s. They say:
\begin{quote}
\begin{small}Here, we assume $a_{i}^{j}=a_{i}^{j'}$ for $j\not=j'$
for any measurement $i$ in two different contexts $j$ and $j'$
(p. 4).\end{small}
\end{quote}
At first they do not claim that this assumption is shared by CbD.
They even correctly state the opposite:
\begin{quote}
\begin{small}However, this argument seems not to be matched by the
CbD approach since the equality $a_{i}^{j}=a_{i}^{j'}$ violates the
double indexing assumption (p. 4).\end{small}
\end{quote}
Stated more precisely, $a_{i}^{j}=a_{i}^{j'}$ contradicts CbD. However,
a few lines later {[}AKB{]} makes a bold promise:
\begin{quote}
\begin{small}We will show that this {[}CbD{]} approach {[}...{]}
also requires the equality $a_{i}^{j}=a_{i}^{j'}$ (p. 4).\end{small}
\end{quote}
This sounds ominous for CbD: if it \emph{requires} something that
it deems meaningless, then it is bad indeed. However, the promised
demonstration consists in once again loosely replicating the \emph{reductio}
reasoning with $a_{i}^{j}$'s in place of $S_{i}^{j}$'s (i.e., with
the built-in ``assumption'' that $a_{i}^{j}=a_{i}^{j'}$), and then
saying
\begin{quote}
\begin{small}This proof considers an equality between two random
variables of each connection: $a_{i}^{j}=a_{i}^{j'}$. (p. 4)\end{small}
\end{quote}
Where, one should ask, is it shown that this equality is required
in the CbD analysis? Apparently, ``showing'' in {[}AKM{]} means
repeating with no justification something its authors promised to
``show.''

\subsection*{Method 2: Criticize a generalization for specializing in some cases
to what it generalizes}

{[}AKB{]} presents this second criticism as a consequence of the reasoning
just considered, but it deserves to be presented as an argument \emph{sui
generis}:
\begin{quote}
\begin{small}if we ignore the double indexing scenario, we can {[}...{]}
convert the CbD notation to the standard representation of Specker
scenario. This demonstrates that CbD adds extra complexity to the
modelling of scenarios like Specker\textquoteright s, without adding
new insights to contextuality (p.4).\end{small}
\end{quote}
The reason this statement deserves to be separated from the preceding
it Method 1 ``arguments'' is that it does make sense. In the case
of consistently connected systems one indeed can use the so-called
\emph{reduced} (single-indexed) couplings \citep{DzhKujHandbook2016}.
In the case of $\mathcal{C}_{3}$ in (\ref{eq:C3}) it would be a
jointly distributed triple $\left(S_{1},S_{2},S_{3}\right)$. Moreover,
if one is willing to overlook certain logical difficulties \citep{DzhKujFoundations2017,Dzh2019},
one can use single-indexed notation for the original random variables,
conveniently confusing them with their reduced coupling. Indeed, traditional
contextuality analysis of $\mathcal{C}_{3}$ would deal with
\begin{equation}
\begin{array}{|c|c|c||c|}
\hline a_{1} & a_{2} &  & c^{1}\\
\hline  & a_{2} & a_{3} & c^{2}\\
\hline a_{1} &  & a_{3} & c^{3}\\
\hline\hline q_{1} & q_{2} & q_{3} & \mathcal{C}_{3}
\\\hline \end{array}\label{eq:single}
\end{equation}
instead of (\ref{eq:C3}). The reason why this single-indexed representation
is strictly speaking incorrect is that in classical probability theory
the existence of any two of the three row-wise joint distributions
in (\ref{eq:single}) implies the existence of a joint distribution
of $\left(a_{1},a_{2},a_{3}\right)$ \citep{DzhKujFoundations2017,Dzh2019}.
In the case of a contextual system, this creates a true paradox. For
instance, if the Bell-type inequality for (\ref{eq:single}) is violated,
i.e., if 
\begin{equation}
\max\left(\begin{array}{c}
\left\langle a_{1}a_{2}\right\rangle +\left\langle a_{2}a_{3}\right\rangle -\left\langle a_{3}a_{1}\right\rangle ,\\
\left\langle a_{1}a_{2}\right\rangle -\left\langle a_{2}a_{3}\right\rangle +\left\langle a_{3}a_{1}\right\rangle ,\\
-\left\langle a_{1}a_{2}\right\rangle +\left\langle a_{2}a_{3}\right\rangle +\left\langle a_{3}a_{1}\right\rangle ,\\
-\left\langle a_{1}a_{2}\right\rangle -\left\langle a_{2}a_{3}\right\rangle -\left\langle a_{3}a_{1}\right\rangle 
\end{array}\right)>1,\label{eq:SZLG}
\end{equation}
we are facing the unresolvable problem of how it is possible that
$\left\langle a_{1}a_{2}\right\rangle ,\left\langle a_{2}a_{3}\right\rangle ,\left\langle a_{3}a_{1}\right\rangle $
are defined (otherwise the inequality cannot be formulated) whereas
$\left\langle a_{1}a_{2}a_{3}\right\rangle $ is not (otherwise the
expression would be $\leq1$). However, the authors who think that
the eristic Method 1 constitutes legitimate reasoning are unlikely
to be bothered by such logical problems. Moreover, single-indexation
of consistently connected systems is a common and convenient practice,
even if logically dubious.

Why, however, this is presented as a deficiency of CbD? Historically,
CbD was proposed as an attempt to understand contextuality in rigorous
probabilisitic terms \citep{DzhKuj2014First}. The fact that it has
rapidly developed into a generalization of the traditional notion
of contextuality to arbitrary, \emph{inconsistently connected} systems
can be viewed as clear evidence that it does provide the ``new insights''
that {[}AKB{]} does not see in it. Inconsistently connected systems
are those in which the distributions of $a_{i}^{j}$ and $a_{i}^{j'}$
for measurements of the same property in two different contexts may
be different. This situation (also designated as ``disturbance''
or ``signaling'') sometimes occurs in quantum physics, and then
CbD is a useful way of dealing with it (see, e.g., \citep{KujDzhLar2015}).
This situation is virtually universal in behavioral applications,
and here CbD serves to correct the researchers who have mistakenly
used the criteria developed in physics for consistently connected
systems \citep{DzhZhaKuj2016,DzhKuj2014}. Some of the authors of
{[}AKB{]} were making this same mistake \citep{Bruzaetal.2009,Kittoetal2012}
until they learned of CbD and began using it to analyze their experiments
\citep{Bruzaetal2015}. As a generalization of the traditional understanding
of contextuality rather than an entirely different notion called by
the same name, CbD must specialize to it (more precisely, to a rigorous
version thereof) for systems that are consistently connected. To criticize
CbD for this is akin to criticizing Lebesgue integration for being
unnecessary when one confines one's attention to Riemann-integrable
functions.

\subsection*{Method 3: Critisize a theory for not addressing issues that are not
in its language}

Another angle of criticism is formulated in the statement that {[}AKB{]}
has
\begin{quote}
\begin{small}demonstrated that there is no clear relation between
the double indexing notation and ontic states (p. 5).\end{small}
\end{quote}
``Demonstrating'' and ``showing'' for the authors of {[}AKB{]}
seems to mean the same as ``stating.'' It is difficult to find a
line of reasoning leading them to their conclusion. Nor is it clear
what they mean by the ontic states. In another place {[}AKB{]} says
that CbD
\begin{quote}
\begin{small}fails to provide a more specific definition of precisely
how the contexts of the double indexed random variables relate to
$\lambda$ (p. 3).\end{small}
\end{quote}
The symbol $\lambda$ in the statement denotes a ``hidden variable,''
which suggests that this notion possibly relates to that of ontic
state. A hidden variable means a random variable of which all random
variables within a given context are functions. That such a variable
exists is an incontrovertible mathematical fact in classical probability
theory, proved as a theorem immediately following from the definition
of a joint distribution \citep{DzhKujHandbook2016,DzhKuj2010}. For
instance, since $a_{1}^{1}$ and $a_{2}^{1}$ in (\ref{eq:C3}) are
jointly distributed, there is a random variable $\lambda$ and two
measurable functions such that $a_{1}^{1}=f\left(\lambda\right)$
and $a_{2}^{1}=g\left(\lambda\right)$. Obviously, $\lambda$ is not
definable uniquely, but one can always define $\lambda$ in the most
economic way, so that if $a_{1}^{1}$ and $a_{2}^{1}$ are also functions
of some random variable $\kappa$, then $\lambda=h\left(\kappa\right)$
for some function $h$ \citep{Dzh2019,DzhKujFoundations2017}. So
the relationship between the observed and hidden variables is far
from being unclear, it is rather straightforward.

It is possible, however, that the authors of {[}AKB{]} understand
ontic states as quantum states. This interpretation is supported by
the various allusions in {[}AKB{]} to how CbD does not address certain
quantum-mechanical notions, such as
\begin{quote}
\begin{small}CbD notation does not discuss scenarios where measurements
are non-orthogonal (p. 4).\end{small}
\end{quote}
One may let it pass that notations rarely ``discuss'' anything.
But one should not let pass that {[}AKB{]} here confuses different
languages and levels of analyzing contextuality. CbD is a theory of
systems of random variables, and as such it does not use the language
of quantum observables and quantum states. But the responses generated
by states and observables are random variables, and at this level
CbD applies to them. CbD does not address quantum-mechanical terms
in the same way it does not address states of mind or computer databases
(but may apply to them if the recorded outcomes are presented as systems
of random variables).

\subsection*{Conclusion}

Unfortunately, not all readers who may be curious to look into {[}AKB{]}
should have sufficient involvement in the area of contextuality to
see all the misrepresentations and glitches in logic this paper contains.
This justifies the need for the present clarifications. I only responded
to the three main lines of criticism found in {[}AKB{]}, represented
by the three eristic methods its authors employ. However, {[}AKB{]}
contains many other, less prominent but no less unfortunate imprecisions
in language and logic. For instance, the authors cite a CbD work as
stating that couplings, like (\ref{eq:coupling}) with respect to
(\ref{eq:C3}), have no empirical meaning, and present this fact as
an obvious deficiency of CbD. Apparently they do not distinguish lack
of empirical meaning (indicating that a notion is a mathematical abstraction
with no empirical referent) from being meaningless (indicating that
a notion is to be discarded). It would take another page or two to
enumerate all such ``pricks and bites'' in {[}AKB{]}. Hopefully
this is not necessary.
\begin{acknowledgments}
I thank Janne Kujala and V\'ictor Cervantes for critically reading
and commenting on this paper.
\end{acknowledgments}


\begin{thebibliography}{99}
\bibitem{AKB}M. Aliakbarzadeh, K. Kitto, and P. D. Bruza, \emph{Is
contextuality about the identity of random variables?} arXiv:1909.13048v1
(2019).

\bibitem{DzhCerKuj2017}E. N. Dzhafarov, V. H. Cervantes, and J. V.
Kujala, \emph{ Contextuality in canonical systems of random variables},
Phil. Trans. Roy. Soc. A 375, 20160389 (2017).

\bibitem{DzhKujFoundations2017}E. N. Dzhafarov and J. V. Kujala,
\emph{Probabilistic foundations of contextuality}, Fortsch. Phys.
- Prog. Phys. 65, 1600040 (1-11) (2017).

\bibitem{KujDzhMeasures}J. V. Kujala and E. N. Dzhafarov, \emph{ Measures
of contextuality and noncontextuality}, Phil. Trans. Roy. Soc. A 377,
20190149 (2019).

\bibitem{LiangSpekkensWiseman2011} Y.-C. Liang, R. W. Spekkens, H.
M. Wiseman, \emph{Specker\textquoteright s parable of the overprotective
seer: A road to contextuality, nonlocality and complementarity.} Phys.
Rep. 506: 1-39 (2011).

\bibitem{DzhKujLar2015}E. N. Dzhafarov, J. V. Kujala, and J.-Å. Larsson,
\emph{ Contextuality in three types of quantum-mechanical systems},
Found. Phys. 7, 762 (2015).

\bibitem{KujDzhLar2015}J. V. Kujala, E. N. Dzhafarov, and J.-Å. Larsson,
\emph{Necessary and sufficient conditions for extended noncontextuality
in a broad class of quantum mechanical systems}, Phys. Rev. Lett.
115, 150401 (2015).

\bibitem{deBarrosetal.2016}J. A. deBarros, J. V. Kujala, and G. Oas,
\emph{Negative probabilities and contextuality}, J. Math. Psych. 74,
34-45 (2016).

\bibitem{Araujoetal2013}M. Araújo, M. T. Quintino, C. Budroni, M.
T. Cunha, and A. Cabello, \emph{All noncontextuality inequalities
for the n-cycle scenario}, Phys. Rev. A 88, 022118 (2013).

\bibitem{KujDzhProof2016}J. V. Kujala and E. N. Dzhafarov, \emph{ Proof
of a conjecture on contextuality in cyclic systems with binary variables},
Found. Phys. 46, 282 (2016).

\bibitem{DzhKujHandbook2016}E. N. Dzhafarov and J. V. Kujala, \emph{Probability,
random variables, and selectivity}. In W. Batchelder, H. Colonius,
E.N. Dzhafarov, J. Myung (Eds), \emph{New Handbook of Mathematical
Psychology}, vol. 1, pp. 85-150. Cambridge University Press (2016).

\bibitem{DzhKuj2010}E. N. Dzhafarov and J. V. Kujala, \emph{The Joint
Distribution Criterion and the Distance Tests for selective probabilistic
causality}. Front. Psych. 1:151 doi: 10.3389/fpsyg.2010.00151 (2010).

\bibitem{Dzh2019}E.N. Dzhafarov, \emph{On joint distributions, counterfactual
values, and hidden variables in understanding contextuality}. Phil.
Trans. Roy. Soc. A 377:20190144 (2019).

\bibitem{DzhKuj2014First}E. N. Dzhafarov and J. V. Kujala, \emph{Contextuality
is about identity of random variables}. Phys. Scrip. T163, 014009
(2014).

\bibitem{DzhZhaKuj2016}E. N. Dzhafarov, R. Zhang, and J. V. Kujala,
\emph{Is there contextuality in behavioral and social systems?} Phil.
Trans. Roy. Soc. A 374, 20150099 (2016).

\bibitem{DzhKuj2014}E. N. Dzhafarov and J. V. Kujala\emph{, On selective
influences, marginal selectivity, and Bell/CHSH inequalities}. Top.
Cog. Sci. 6, 121-128 (2014).

\bibitem{Bruzaetal.2009}P. D. Bruza, K. Kitto, D. Nelson, and C.
McEvoy, \emph{Is there something quantum-like about the human mental
lexicon}? J. Math. Psych., 53, 362-377 (2009).

\bibitem{Kittoetal2012}K. Kitto and P. D. Bruza, \emph{Tests and
Models of Non-compositional Concepts}. In Miyake, N, Peebles, D.,
\& Cooper, R.P. (Eds.) Proceedings of the 34th Annual Conference of
the Cognitive Science Society, Cognitive Science Society, Sapporo,
Japan (2012).

\bibitem{Bruzaetal2015}P. D. Bruza, K. Kitto, B. J. Ramm, and L.
Sitbonc. \emph{A probabilistic framework for analysing the compositionality
of conceptual combinations}, J. Math. Psych., 67, 26-38 (2015).
\end{thebibliography}
\end{document}